\documentclass[12pt]{article}

\usepackage{a4}
\usepackage{a4wide}

\usepackage{amsmath}
\usepackage{amscd}
\usepackage{amsfonts}
\usepackage{amssymb}
\usepackage{mathrsfs}
\usepackage{fancybox} 
\usepackage{enumerate}
\usepackage{accents} 

\usepackage{bm}
\usepackage{amscd}
\usepackage{graphicx}
\usepackage[dvips]{color}
\usepackage{epsfig}

\makeatletter

\@addtoreset{equation}{section}
\makeatother

\usepackage{amsthm} 
\usepackage{ascmac}  

\def\half{{1\over2}}

\def\={\stackrel{\bullet}{=}}

\def\({\left(}
\def\){\right)}
\def\[{\left[}
\def\]{\right]}

\def\cO{{\cal O}}

\def \be {\begin{equation}}
\def \ee {\end{equation}}
\def \beqa {\begin{eqnarray}}
\def \eeqa {\end{eqnarray}}
\def \beal#1 {\begin{align}#1\end{align}}
\def \bes#1 {\begin{equation}\begin{split}#1\end{split}\end{equation}}
\def \nn {\notag\\}

\def\aver#1{\left\langle #1 \right\rangle}

\makeatletter

\@addtoreset{equation}{section}
\makeatother

\begin{document}

\begin{titlepage}
\title{
\vspace{-2cm}
\begin{flushright}
\normalsize{ 
YITP-17-72}
\end{flushright}
       \vspace{1.5cm}
Holographic computation of quantum corrections to the bulk cosmological constant
\vspace{1.5cm}
}
\author{
Sinya Aoki\thanks{saoki[at]yukawa.kyoto-u.ac.jp
},\; Janos Balog\thanks{balog.janos@wigner.mta.hu},\; Shuichi Yokoyama\thanks{shuichi.yokoyama[at]yukawa.kyoto-u.ac.jp
}
\\[25pt] 
${}^{*}{}^{\ddagger}$ {\it Center for Gravitational Physics,} \\
{\it Yukawa Institute for Theoretical Physics, Kyoto University,}\\
{\it Kitashirakawa-Oiwakecho, Sakyo-Ku, Kyoto, Japan}
\\[10pt]
${}^\dagger$ {\it Institute for Particle and Nuclear Physics,}\\ 
{\it Wigner Research Centre for Physics,}\\ 
{\it MTA Lend\"ulet Holographic QFT Group,}\\
{\it 1525 Budapest 114, P.O.B.\ 49, Hungary}
\\[10pt]
}

\date{}
\maketitle

\thispagestyle{empty}


\begin{abstract}
\vspace{0.3cm}
\normalsize
We explore the program of the construction of the dual bulk theory in the flow equation approach. 
We compute the vacuum expectation value of the Einstein operator at the next to leading order in the $1/n$ expansion using a free O($n$) vector model. 
We interpret the next to leading correction as the quantum correction to the cosmological constant of the AdS space. 
We finally comment on how to generalize this computation to matrix elements of the Einstein operator for excited states.
\end{abstract}
\end{titlepage}

\section{Introduction} 
\label{Intro} 

It has passed two decades since the AdS/CFT correspondence was discovered \cite{Maldacena:1997re} (see \cite{Aharony:1999ti,Klebanov:2000me,DHoker:2002nbb} for reviews).  
This is a theoretical realization of holography \cite{tHooft:1993dmi,Susskind:1994vu} and provides an alternative framework to explore strongly coupled gauge theory or quantum gravity from its dual weakly coupled theory from a computable relation of observables \cite{Gubser:1998bc,Witten:1998qj}.
Even though it is very difficult to give a complete proof for the AdS/CFT correspondence, as is usual with a strong-weak duality, there have been several attempts to deepen the understanding of the duality by extending a boundary conformal field theory (CFT) to a bulk gravitational one \cite{Banks:1998dd,Balasubramanian:1998sn,Balasubramanian:1998de} (see also \cite{Balasubramanian:1999ri,Bena:1999jv,Lifschytz:2000bj}). 

One of the recent focuses in the study of the AdS/CFT correspondence is on how diffeomorphism invariance is encoded in a boundary CFT and the Einstein equation is reproduced from boundary data. 
Such a study was initiated at the linear order of perturbation around the AdS background by using the entanglement entropy \cite{Nozaki:2013vta,Bhattacharya:2013bna,Faulkner:2013ica}.
A recent study in \cite{Faulkner:2017tkh} elegantly reproduced the Einstein equation with a fixed gauge at the second order by incorporating a geometrical identity \cite{Hollands:2012sf} (see also \cite{Sarosi:2017rsq}). 
In the holographic renormalization group approach by using the local renormalization group \cite{Osborn:1991gm} (see \cite{Imbimbo:1999bj,Papadimitriou:2004ap,Lee:2012xba,Lee:2013dln} and also \cite{Skenderis:2002wp} for a review and references therein), it was shown in an abstract way that the bulk diffeomorphism invariance is fully encoded in the form of the Poisson algebra of the RG Hamiltonian by its Wess-Zumino consistency condition 
\cite{Shyam:2016zuk,Shyam:2017qlr}. 

In this situation this paper aims at proposing a new scheme to compute bulk dynamical observables from a boundary CFT by employing a new approach of the AdS/CFT incorporating a flow equation \cite{Aoki:2015dla,Aoki:2016ohw,Aoki:2017bru,Aoki:2017uce}, which was introduced to smear operators so as to resolve the UV divergence arising in the coincidence limit \cite{Albanese:1987ds,Narayanan:2006rf,Luscher:2010iy}. 
One of the virtues in the flow equation approach is to access a dual geometry directly even in a non-conformal case, which emerges after the metric operator condenses \cite{Aoki:2015dla,Aoki:2016ohw}. 
This enables one to elucidate that the boundary conformal transformation converts to the bulk AdS isometry precisely after taking into account the quantum effect at the boundary \cite{Aoki:2017bru} (see also \cite{Jevicki:1998qs,Das:2003vw}), and to provide an AdS geometry whose boundary is a general conformally flat manifold \cite{Aoki:2017uce}.  
In this paper, we further pursue this direction, in order to compute
the quantum correction to the cosmological constant of the bulk AdS space through the vacuum expectation value of the Einstein tensor at the next to leading order of the $1/n$ expansion.
This may provide a deeper understanding of the bulk dynamics in the proposal.

The rest of this paper is outlined as follows. 
In Sec.~\ref{sec:pre-geometry}, we explain our approach to define bulk operators from a boundary CFT containing a scalar primary operator though the flow equation. We introduce various ``quantum operators''  here. In Sec.~\ref{sec:Einstein}, we present our results. We first show that the vacuum expectation value of the metric operator describes the bulk AdS space at the leading order of the $1/n$ expansion for the free massless $O(n)$ vector model. We then calculate the next to leading order corrections and show that, while the metric operator receives no corrections, the Einstein tensor has
$1/n$ correction proportional to the metric, which can be interpreted as a quantum correction to the cosmological constant of the AdS space. 
In Sec.~\ref{sec:symmetry}, we explain why the quantum correction to the Einstein tensor is proportional to the metric, using the conformal symmetry of the original $d$ dimensional theory.
In Sec.~\ref{sec:discussion} we summarize our results and discuss some extensions in the future.
Some technical details are given in appendices.
In appendix~\ref{app:CovariantPerturbation}, we set up a covariant formulation for the $1/n$ expansion around the background metric. In appendix~\ref{app:2pt}, we calculate various 2-point functions for the metric operator around its vacuum expectation value, which are necessary for the $1/n$ expansion.

\section{Pre-geometric operators} 
\label{sec:pre-geometry}

In this section we propose the method to compute dual observables in the flow equation approach. 
In this approach we generally construct $d$-dimensional operators parametrized by a flow time which become seeds of geometric objects in the bulk theory.  
We refer to such operators as pre-geometric operators. 
Then physical observables in the bulk  are obtained by taking the vacuum expectation value of the pre-geometric operators.   

\subsection{Metric operator} 

In this subsection we illustrate how to construct a metric operator in the flow equation approach. See also \cite{Aoki:2017bru} for more details.
For this purpose as well as later convenience, we consider a $d$-dimensional quantum field theory whose elementary fields are $n$ real scalar fields denoted by $\varphi^a(x)$ with $a=1,2,\cdots, n$. 
Then we smear the elementary fields so as to remove the short distance singularity, which is described by a flow equation of a generic form given as
\beqa
\frac{\partial \phi^a(x;t)}{\partial {t}} &=& - \left.\frac{\delta S_f(\varphi)}{\delta \varphi^a(x)}\right\vert_{\varphi(x)\to\phi(x;t)},\quad \phi^a(x;0)=\varphi^a(x). 
\eeqa
Here $S_f$ is a smearing functional describing how to smear operators, which is in principle independent of the action of the original theory controling the dynamics of $\varphi$. This defines a procedure to smear the original field $\varphi^a$ into a smeared field $\phi^a$ like the block spin transformation of $\varphi^a$ in a non-local fashion,
and $\phi^a$ is called the flow field corresponding to $\varphi^a$.
The dynamics generated by the flowed operators $\phi^a(x;t)$ constitutes a holographic theory with the smearing scale $\sqrt{t}$ as the holographic direction. 

To approach the dual AdS geometry from CFT, it is convenient to consider the free flow, which is realized by choosing $S_f$ as the free action:
\beqa
\frac{\partial \phi^a(x;t)}{\partial {t}} &=& (\partial^2 -m^2) \phi^a(x;t). 
\eeqa
The solution is given by
\beqa
\phi^a(x;t) &=& e^{t(\partial^2-m^2)}\varphi^a(x) = \frac{e^{-tm^2}}{(4\pi t)^{d/2}} \int d^d y\, e^{-(x-y)^2/4t} \varphi(y).
\eeqa
The free flow equation is formally the same form as the heat equation, so that the smeared operator is given by superposing all the original operators inserted at each point of the space with the gaussian weight whose standard deviation is the smearing scale $\sqrt t$. 

Following the standard renormalization group transformation procedure, we (re)normalize the smeared field $\phi^a$  as
\beqa
\sigma^a(x;t) := \frac{\phi^a(x;t)}{\sqrt{\langle \sum_{a=1}^n \phi^a (x;t)^2 \rangle}} ,
\eeqa
where $\langle O(\varphi) \rangle$ denotes the quantum average with the original $d$ dimensional action $S$ as 
\beqa
\langle O(\varphi) \rangle := \frac{1}{Z} \int {\cal D}\varphi\, O(\varphi)\, e^{-S(\varphi)}, \
Z = \int {\cal D}\varphi\,  e^{-S(\varphi)}.
\eeqa
Note that this operator is well-defined due to the fact that the flowed operators are free from the contact singularity. 

We can introduce the metric operator, which becomes the metric in the holographic space after taking the quantum average, as 
\beqa
\hat g_{MN} (x;t) := L^2 \sum_{a=1}^n {\partial \sigma^a(x;t) \over \partial z^M} {\partial \sigma^a(x;t) \over \partial z^N}, 
\eeqa
where  $L$ is a constant with  the dimension of length, and
$z^M=(x^\mu, \tau)$ with $\tau=\sqrt{2 d t}$, which will be regarded as the $d+1$ dimensional coordinates.
The vacuum expectation value (VEV) of the metric is called the induced metric, which is denoted by $g_{MN}(z) := \langle \hat g_{MN}(x;t) \rangle$. 

It was shown in \cite{Aoki:2017bru} that the induced metric $g_{MN}(z)$ 
becomes a quantum information metric called the Bures or Helstrom metric for the Hilbert space generated by the flowed fields. 
This holds for a general (even non-conformal) quantum field theory. 

\subsection{Other pre-geometric operators} 

Once the metric operator is constructed, other pre-geometric operators are defined by replacing the metric which appears in the definition of the corresponding (classical) geometric object with the metric operator. 
For example, the Levi-Civita connection operator is defined by  
\beal{
\hat\Gamma^{ M }_{L N }(x;t) =\half \hat g^{ M  P }(x;t) (\hat g_{ P \{ N , L \}}(x;t) -\hat g_{ N  L , P }(x;t) )
}
where $X_{\{x,y\}}:= X_{x,y} + X_{y,x}$. 
Curvature operators are defined by 
\beal{
\hat R_{ L  P }{}^{ M }{}_{ N }(x;t)=&\partial_{[ L }\hat \Gamma^{ M }_{ P ] N }(x;t)+\hat\Gamma_{[ LQ }^{ M }(x;t)\hat\Gamma_{ P ]N}^{ Q }(x;t) \\
\hat R_{ P  N }(x;t)=&\hat R_{ M  P }{}^{ M }{}_{ N }(x;t), \\
\hat R(x;t)=&\hat g^{ P  N }(x;t)\hat R_{ P  N }(x;t), 
}
where $X_{[x,y]}:= X_{x,y} -X_{y,x}$. 
Finally the Einstein tensor operator is defined by 
\be 
\hat G_{MN}(x;t) = \hat R_{MN}(x;t) - \half \hat g_{MN}(x;t) \hat R(x;t) .
\ee

\section{Induced Einstein tensor and bulk interpretation}
\label{sec:Einstein} 

In this section, we evaluate the VEV of the Einstein tensor operator for a free $O(n)$ vector model at the next-to-leading order (NLO) in the $1/n$ expansion.
We then interpret the induced Einstein tensor as the bulk stress-energy tensor through the Einstein equation as
\beqa 
\langle \hat G_{AB}\rangle &=& T^{\rm bulk}_{AB}.
\label{interpretation} 
\eeqa
Since the Einstein tensor is now evaluated on the vacuum, it is natural to think that the corresponding bulk stress-energy tensor consists only of the cosmological constant term:
\beqa
T^{\rm bulk}_{AB} &=& -\Lambda g_{AB}.
\label{interpretationVacuum}
\eeqa
In what follows, we compute the cosmological constant $\Lambda$ at the NLO in the $1/n$ expansion.

\subsection{Leading order}

Let us first compute induced geometric observables for a free $O(n)$ vector model at the leading order (LO) in the $1/n$ expansion. 
For this computation we can use the result in \cite{Aoki:2017bru}, where we computed the induced metric for an arbitrary CFT which contains a real scalar primary operator $\varphi(x)$ of a general conformal dimension $\Delta$.
The two-point function of the normalized field $\sigma$ is  
\beqa
\langle \sigma^a(x;t) \sigma^b(y;s) \rangle_{\rm CFT} 
=\frac{\delta^{ab}}{n}\left(\frac{2\sqrt{ts}}{t+s}\right)^\Delta F\left(\frac{(x-y)^2}{{t+s}}\right),
\label{2ptsigma}
\eeqa
where 
$F(0)=1$ and $2 d F^\prime(0) = -\Delta$.
Explicitly, $F(x)$ is given by
\beqa
F(x) = \frac{\Gamma(d/2)}{\Gamma(\Delta)\Gamma(d/2-\Delta)}&\displaystyle \int_0^1& dv\,  v^{\Delta-1}(1-v)^{d/2-\Delta-1} e^{-x v/4}.  
\eeqa
In the current case, $\Delta = (d-2)/2$. 

The VEV of the induced metric becomes 
\beqa
g_{AB} &=&  \frac{L^2\Delta}{\tau^2}\delta_{AB}.
\label{eq:AdS_metric}
\eeqa
Using the factorization in the large $n$ limit as $\langle \hat g_{AB}\hat g_{CD}  \rangle =g_{AB} g_{CD} $ etc. we then obtain
the induced Levi-Civita connection at the LO as
\beqa
\Gamma^A_{BC} &=& -\frac{1}{\tau} \rho^{A\tau}_{BC}, \qquad
\rho^{A\tau}_{BC} :=\delta_{B\tau}\delta_{AC} + \delta_{C\tau}\delta_{AB} -\delta_{A\tau}\delta_{BC}, \nn 
\Gamma^A_{BC,D} &=& \frac{1}{\tau^2} \rho^{A\tau,D}_{BC},
\qquad
\rho^{A\tau,D}_{BC} :=\delta_{BD}^\tau\delta_{AC} + \delta_{CD}^\tau\delta_{AB} -\delta_{AD}^\tau\delta_{BC}, \ \delta^\tau_{AB} :=\delta_{A\tau}\delta_{B\tau}.
\eeqa
Induced curvatures at the LO are computed as 
\beqa
R_{ A  B }{}^{ C }{}_{ D }&=&
- { g^C_{[A} g_{B]D} \over L^2\Delta}, \quad 
R_{AB} = -\frac{d}{L^2\Delta} g_{AB}, \quad 
R= -\frac{d(d+1)}{L^2\Delta},
\eeqa
which determine the VEV of the Einstein tensor at the LO as 
\beqa
G_{AB} &=&\frac{d(d-1)}{2L^2\Delta} g_{AB}. 
\eeqa
Therefore the cosmological constant of the dual geometry is evaluated through \eqref{interpretationVacuum} at the LO as 
\beqa
\Lambda= -\frac{d(d-1)}{2L^2\Delta} + \cO\left({1\over n}\right), \qquad 
\eeqa
which implies that the AdS radius of the dual geometry is given at LO by $L_{\rm AdS} ^2 = L^2\Delta + \cO(1/n)$. 

\subsection{Next to leading order}

We proceed to the next to leading order computation of the induced Einstein tensor. 
For this purpose we employ a covariant perturbation expansion: 
that is, we expand an arbitrary operator $\hat X$ built from the metric operator $\hat g_{AB}$ around the vacuum by plugging $\hat g_{AB} = g_{AB} + \hat h_{AB}$ and expanding in powers of the fluctuation field $\hat h_{AB}$.
We indicate terms in this expansion with a certain power of $\hat h_{AB}$ with the corresponding number of dots like 
\be 
\hat X = X + \dot X + \ddot X + \cdots . 
\ee
Terms with increasing number of $\hat h_{AB}$ fields are more and more suppressed in the $1/n$ expansion.
Then the VEV of the operator $\hat X$ reduces to that of correlation functions of the fluctuation field $\hat h$. 
In this notation the Einstein tensor at the NLO is given by
\beqa
\langle \ddot G_{AB}\rangle &=& 
\langle \ddot R_{AB}\rangle 
-\frac{1}{2} \langle \hat h_{AB} \dot R \rangle -\frac{1}{2} g_{AB} \langle \ddot R\rangle.
\eeqa
We summarize results of this expansion in Appendix \ref{app:CovariantPerturbation}. 

For a free $O(n)$ vector model, {the result \eqref{eq:AdS_metric} does not receive any correction}. 
Using results for 2-point functions of $\hat h_{AB}$, which are calculated in appendix~\ref{app:2pt},
we obtain
\beqa
\langle \ddot R_{AB}\rangle &=& \frac{1}{n(d+2)\tau^2}\left( \chi_{ABCC} -\chi_{ACBC}\right)
=  -\frac{d}{n\tau^2} \delta_{AB},
\\
-\frac{1}{2}\langle \hat h_{AB} \dot R\rangle &=& \frac{1}{n(d+2)\tau^2}\left(\chi_{ABCC} -\chi_{ACBC}\right) =-\frac{2d}{n\tau^2} \delta_{AB},
\eeqa
and
\beqa
g^{AB} \langle \ddot R_{AB}\rangle &=& \frac{1}{n(d+2)L^2\Delta}\left( \chi_{CCDD} -\chi_{CDCD}\right)= \langle \hat h^{AB} \dot R_{AB}\rangle, \\
\langle \hat h^{AC} \hat h_{C}^{\ \ B}\rangle R_{AB} &=& -\frac{d(d+1)(d+2)}{n L^2\Delta},
\eeqa
so that we have
\beqa
-\frac{1}{2} g_{AB} \langle \ddot R \rangle &=& \frac{d(d+1)(d+2)}{2n\tau^2}\delta_{AB},
\eeqa
where we use
\beqa
\chi_{ABCC} &=& (\Delta +1)\{(d+5)\delta_{AB} +2(d+2)\delta^\tau_{AB}\} -(d+1)(d+2)\delta_{AB}, \\
\chi_{ACBC} &=& (\Delta +1)\{(d+5)\delta_{AB} +2(d+2)\delta^\tau_{AB}\} -(d+2)\delta_{AB}, \\
\chi_{CDCD} &=&  (\Delta +1)\{(d+5)(d+1) +2(d+2)\} -(d+1)(d+2),
\eeqa
and the definition of $\chi_{ABCD}$ can be found in appendix~\ref{app:2pt}.
Finally we obtain%
\footnote{ 
This result was checked in a slightly different computation by 
\beal{
\aver{\ddot G_{MN}}
=& \aver{\ddot R_{PQ}}(\delta^P_M \delta^Q_N -\half g_{MN} g^{PQ}   ) +\half(- g^{PQ}\delta^{R}_M \delta^{S}_N + g_{MN} g^{RP} g^{SQ} )\aver{\hat h_{RS}  \dot R_{PQ} } \nn
&+\half  \aver{\hat h_{MN} \hat h^{PQ} }R_{PQ} -\half g_{MN} \aver{ \hat h^{ Q  P }\hat h_{ P }^R } R_{ Q R}, 
}
where  $\aver{\hat h_{RS}  \dot R_{PQ} } $ is evaluated as
$\aver{\hat h_{RS}  \dot R_{PQ} } =  \dfrac{- g_{\{RQ} g_{S\}P}+ 2 g_{RS} g _{PQ}}{ n \Delta}.$
}
\beqa
\langle \ddot G_{AB}\rangle
=\frac{(d-1)d(d+4)}{2 n L^2\Delta} g_{AB}. 
\eeqa
This final result is manifestly covariant even though the calculation in the intermediate step contains non-covariant terms. 
This is a non-trivial check of our result. 


As a result the induced Einstein tensor evaluated at the vacuum is given by 
\be 
\langle \hat G_{AB}\rangle =\frac{d(d-1)}{2L^2\Delta}g_{AB}\(1 + \frac{d+4}{n }\) + \cO\({1\over n^{2}}\). 
\ee
As asserted at the beginning of this section, this quantity is related to the bulk stress energy tensor through the bulk Einstein equation \eqref{interpretation}. 
Through \eqref{interpretationVacuum} the cosmological constant and the AdS radius are determined at the NLO as 
\beal{ 
\Lambda=-\frac{d(d-1)}{2L^2\Delta}\left(1 + \frac{d+4}{ n }\right) + \cO\({1\over n^{2}}\), \quad 
L_{\rm AdS}^2=L^2\Delta\left(1 - \frac{d+4}{n }\right)+\cO\({1\over n^{2}}\).  
}
Since $ \langle \hat g_{AB}\rangle$ has no NLO corrections and thus has the same classical relation to the Einstein tensor at the LO, 
the $1/n$ correction to the cosmological constant comes purely from the quantum effect to the Einstein tensor {of the dual gravity theory to the O($n$) free vector model, which is conjectured as the free higher spin theory on AdS$_{d+1}$ \cite{Klebanov:2002ja}}. 

\section{Symmetry constraints}
\label{sec:symmetry}

Finally we discuss our result in terms of the conformal symmetry or AdS isometry.
In the previous publications \cite{Aoki:2017bru,Aoki:2017uce}, we showed that the conformal transformation converts to the AdS isometry after taking the quantum average. 
In what follows, we show that pre-geometric operators computed at the $1/n$ level in the previous sections become covariant under the symmetry after taking the quantum average.

For this purpose, as done in \cite{Aoki:2017bru,Aoki:2017uce}, we divide the infinitesimal conformal transformation for the $\sigma(x;t)$ as
\beqa
\delta^{\rm conf} \sigma(x;t) &=& \delta^{\rm diff}\sigma(x;t) + \delta^{\rm extra}\sigma(x;t),
\label{Separation} 
\eeqa 
where $ \delta^{\rm diff}$ generates the isometry of the AdS space while $\delta^{\rm extra}$ is the remaining contribution, which was shown to vanish after taking the vacuum expectation value for the metric operator.
Then in order to show that a pre-geometric operator $\hat T$ behaves  in a covariant manner under the AdS isometry, we need to show that $\langle \delta^{\rm diff}\hat T \rangle = 0,$ which is equivalent to $\langle \delta^{\rm extra} \hat T \rangle = 0$.

To this end, the explicit expression of $\delta^{\rm diff}\sigma(x;t), \delta^{\rm extra}\sigma(x;t)$ is not necessary, but we only need the operation of $\delta^{\rm extra}$ on the 2-pt function, which is given by
\beqa
\langle \delta^{\rm extra}\{ \sigma(x_1;t_1)\sigma(x_2;t_2) \} \rangle &=& -16 d \left({2\tau_1\tau_2\over \tau_1^2+\tau_2^2}\right)^\Delta {(\tau_1^2-\tau_2^2)\over (\tau_1^2+\tau_2^2)^2}
b\cdot(x_1-x_2) (x_1-x_2)^2 \nn 
&\times& 
F^{\prime\prime}\left(\frac{2d(x_1-x_2)^2}{\tau_1^2+\tau_2^2}\right).
\eeqa
Here $b_\mu$ is the parameter of an infinitesimal conformal transformation. Using this, we have
\beqa
\langle {\delta^{\rm extra}\{ \hat h_{AB;E} \hat h_{CD;F} \} }\rangle &=& - C(\tau) b_\mu 
\left[\delta_{BD}\left\{ \delta_{E\tau} \rho^{d}_{ACF\mu}  + \delta_{F\tau} \rho^{d}_{ACE\mu} 
+ \delta_{A\tau} \rho^{d}_{CEF\mu} + \delta_{C\tau} \rho^{d}_{AEF\mu}\right\}\right.    \nn 
&+&\left. (C\leftrightarrow D) + (A\leftrightarrow B) + \mbox{both} \right],
\label{eq:extra}
\eeqa
where
\beqa
C(\tau)&:=& \frac{16 d L^4\Delta F^{\prime\prime}(0)}{n \tau^5}, \quad
\rho^{d}_{ABC\mu} :=\delta_{AB}^{\tau}\delta_{C\mu} + \delta_{AC}^{\tau}\delta_{B\mu}+\delta_{BC}^{\tau}\delta_{A\mu}.
\eeqa

We need to consider
\beqa
\langle \ddot R_{AB}\rangle &=& \frac{\tau^4}{4 L^4\Delta^2}\left[ 
\langle  \hat h_{CC;D}(\hat h_{DA;B} +\hat h_{DB;A} -\hat h_{BA;D} ) \rangle
+2  \langle  \hat h_{AC;D} \hat h_{BD;C} -\hat h_{AD;C}  \hat h_{BD;C} ) \rangle\right. \nn 
&+&\left. \langle \hat h_{CD;A} \hat h_{CD;B} \rangle \right], \\
\langle \hat h_{AB}\dot R\rangle &=& \frac{\tau^4}{L^4\Delta^2}
\langle \hat h_{AB;C} (\hat h_{DD;C} - \hat h_{CD;D})\rangle 
+ {d \tau^2 \over L^4\Delta^2} \langle \hat h_{AB} \hat h_{CC}   \rangle..
\eeqa

Eq.~(\ref{eq:extra}) leads to
\beqa
\langle \delta^{\rm extra} \{ \hat h_{CC;D} \hat h_{DA;B} \} \rangle 
&=& -4(d+2) C(\tau)\left[\delta_{A\tau} b_B+ b_A\delta_{B\tau} \right] \nn 
&=& \langle \delta^{\rm extra} \{\hat h_{CC;D} \hat h_{AB;D} \}\rangle= \langle \delta^{\rm extra} \{ \hat h_{CD;D} \hat h_{AB;C} \}\rangle,  \\
\langle \delta^{\rm extra} \{ \hat h_{AC;D} \hat h_{BD;C} \}\rangle &=& -3(d+2) C(\tau)\left[\delta_{A\tau} b_B
+ b_A\delta_{B\tau} \right] ,  \\
\langle \delta^{\rm extra} \{ \hat h_{AD;C} \hat h_{BD;C} \}\rangle &=& -(d+3)(d+2) C(\tau)\left[\delta_{A\tau} b_B
+ b_A\delta_{B\tau} \right] ,  \\
\langle \delta^{\rm extra} \{ \hat h_{CD;A} \hat h_{CD;B} \}\rangle &=& - 2(d+2)(d+2) C(\tau)\left[\delta_{A\tau} b_B
+ b_A\delta_{B\tau} \right] ,
\eeqa
{where $b_\tau=0$. Then} we obtain
\beqa
\langle \delta^{\rm extra} \ddot R_{AB}\rangle &=& \frac{\tau^4 C(\tau)}{4 L^4\Delta^2}
(d+2)\left[\delta_{A\tau} b_B+ b_A\delta_{B\tau} \right] \nn 
&\times& \left\{ -4 -4 +4 +2(-3+ d+3) -2(d+2)\right\} = 0,  \\
\langle \delta^{\rm extra} \{ \hat h_{AB}\dot R \}\rangle &=&
\frac{\tau^4 C(\tau)}{L^4\Delta^2}
(d+2)\left[\delta_{A\tau} b_B+ b_A\delta_{B\tau} \right] (4-4) = 0.
\eeqa

These guarantee that each term is covariant under isometry, and thus proportional to $g_{AB}$, as seen in the previous section.

\section{Discussion} 
\label{sec:discussion}

In this paper, we have constructed the holographic space from the primary scalar field in a free massless $O(n)$ vector model by a flow equation
at the next to leading order (NLO) in the $1/n$ expansion. 
We investigated some properties of the bulk theory,   
by calculating the induced Einstein tensor in the $1/n$ expansion.
After defining pre-geometric operators, we have calculated the VEV of the Einstein tensor operator at the NLO in the $1/n$ expansion.
{As a result the NLO correction appeared in the VEV of the Einstein operator but not in the induced metric $g_{AB}$.}
We therefore regarded the NLO correction of the induced Einstein operator as the quantum correction to the cosmological constant from the dual gravity theory on AdS space, which is proposed as the free higher spin theory \cite{Klebanov:2002ja}.
We have also shown that the stress-energy tensor for the vacuum state is covariant and proportional to $g_{AB}$, thanks to the conformal symmetry of the boundary theory, which turns into the bulk isometry of the AdS space.    

In our approach, the bulk AdS radial direction emerges as the smearing scale for the boundary CFT.
It is important to clarify how to determine the whole structure of the bulk theory.
The bulk stress-energy tensor corresponding to the vacuum state calculated in this paper may give a hint to construct the bulk theory.

In this paper we computed the 1-loop correction to the cosmological constant in the bulk theory, which is supposed to be
the free higher spin theory \cite{Klebanov:2002ja}, from the dual free O$(n)$ vector model in the proposed framework. On the other hand, one-loop tests of higher-spin/vector model duality were already done in \cite{Giombi:2013fka,Giombi:2014iua}. (See also \cite{Camporesi:1993mz,Giombi:2014yra,Jevicki:2014mfa,Beccaria:2014xda,Bae:2016rgm}.)
Their intriguing result is that the logarithmic divergence in the 1-loop correction to the free energy cancels for higher spin gauge theories employing a standard zeta function regularization in accordance with the analysis of sphere partition functions of their dual vector models. 
In particular it was confirmed that the finite part vanishes for a free higher spin theory with its scalar field obeying the standard boundary condition.  
In the flow equation approach the 1-loop correction to the cosmological constant, which presumably corresponds to that of the vacuum energy, is free from UV divergence without specifying any regularization scheme. 
This is because the computation of the 1-loop correction reduces to the 2-point function of the flowed field, which has no UV divergence by the construction. Our result of the finite part correction to the cosmological constant for a free higher spin theory does not vanish in any dimension. It is highly important to fill the gap between these results, which could be done by identifying a bulk local operator in the flow equation approach. 

The next important step is to evaluate the bulk stress energy tensor corresponding to excited states.
Indeed, we can easily generalize the computation of the VEV for the Einstein operator presented in this paper to that of arbitrary states as follows.
We consider a set of states $\{\vert O \rangle\}$ in CFT with the inner product $ \langle O \vert O^\prime\rangle = \delta_{O,O^\prime}$, where the meaning of $\delta_{O,O^\prime}$ depends on the type of states. 
Then we evaluate the matrix element of the Einstein operator in the $1/n$ expansion by using the covariant perturbation given in appendix \ref{app:CovariantPerturbation} as
\beqa
\langle O \vert \hat G_{AB} \vert O^\prime \rangle &=&
\langle O \vert G_{AB} \vert O^\prime \rangle +
\langle O \vert \dot G_{AB} \vert O^\prime \rangle +
\langle O \vert  \ddot G_{AB}  \vert O^\prime \rangle +\cdots \nn 
&=& \left\{ G_{AB} +\langle 0 \vert  \ddot G_{AB} \vert 0 \rangle \right\}
\delta_{O,O^\prime} +\langle O \vert  \dot G_{AB} \vert O^\prime \rangle +
\langle O \vert  \ddot G_{AB}  \vert O^\prime \rangle_c   + O\left(\frac{1}{n^2}\right). ~~~~~~
\label{matrixElement}
\eeqa
where $\langle O \vert \hat X \vert O^\prime \rangle_c := \langle O \vert \hat X \vert O^\prime \rangle - \langle 0 \vert  \hat X\vert 0 \rangle \delta_{O,O^\prime}$  for an arbitrary operator $\hat X$.
As asserted in section \ref{sec:Einstein}, we interpret the matrix element of the Einstein operator as the bulk stress energy tensor through \eqref{interpretation}, which we may call the quantum Einstein equation.
It is natural to interpret in this way that the corresponding bulk stress energy tensor consists of the cosmological constant and the contribution from the matter field in the bulk: 
\be 
T^{\rm bulk}_{AB} = -\Lambda g_{AB}^{\rm mat} + T^{\rm mat}_{AB}, \qquad
g_{AB}^{\rm mat} :=  \langle O \vert \hat g_{AB} \vert O^\prime \rangle.
\ee
Notice that we have already calculated the 1st term in \eqref{matrixElement} as  
\beqa
G_{AB} + \langle 0 \vert  \ddot G_{AB} \vert 0 \rangle &=& -\Lambda
g_{AB}, \qquad \Lambda= -\frac{d(d-1)}{2L^2\Delta}\left(1 + \frac{d+4}{ n }\right) +\cdots, 
\eeqa
which represents the vacuum contribution.
Therefore the contribution of the matter field to the bulk stress energy tensor is given by 
\beqa
 T^{\rm mat}_{AB}&=&
\langle O \vert \dot G_{AB} \vert O^\prime \rangle + 
\langle O \vert  \ddot G_{AB}  \vert O^\prime \rangle_c
+\Lambda \langle O \vert \hat g_{AB} \vert O^\prime \rangle_c .
\eeqa 
It is very important to compute this bulk stress energy tensor in the construction of the dual bulk theory beyond the vacuum or geometry level. 
We are currently calculating $T^{\rm mat}_{AB}$
, and will report the result elsewhere.

This program can be extended to the case of the 
$\lambda\varphi^4$ theory in 3-dimensions. In the previous investigation \cite{Aoki:2016env},
while the induced metric describes the AdS$_4$ space at the leading order, the next leading order corrections make the space asymptotically AdS
only in the UV and IR limits with different radii.
These two limits correspond to the asymptotically free UV fixed point and the Wilson-Fischer IR fixed point of the boundary theory, respectively.
Then it would be interesting to investigate how the stress-energy tensor in the bulk behaves from UV to IR at the NLO.
We expect that this behaves in a similar manner to the one computed from the corresponding solution for the dual bulk (higher-spin) theory. 
 
We hope to report the progress on these issues in the near future.

\section*{Acknowledgement}
S. A. is supported in part by the Grant-in-Aid of the Japanese Ministry of Education, Sciences and Technology, Sports and Culture (MEXT) for Scientific Research (No. JP16H03978),  
by a priority issue (Elucidation of the fundamental laws and evolution of the universe) to be tackled by using Post ``K" Computer, 
and by Joint Institute for Computational Fundamental Science (JICFuS).
This  work  was partially supported by the Hungarian National Science Fund OTKA (under K116505). 

\appendix
\section{Covariant perturbation}
\label{app:CovariantPerturbation}

We introduce the fluctuation of the metric operator around its VEV as
\beqa
\hat g_{AB} &=& g_{AB} + \hat h_{AB},   
\eeqa
where $g_{AB} =\aver{\hat g_{AB}}$. Note $\hat h_{AB}=\hat h_{BA}$.
The inverse is thus expanded as
\beqa
\hat g^{AB} = g^{AB} - g^{AC} \hat h_{CD} g^{DB} +\cdots = g^{AB} -\hat h^{AB} + \hat h^{AC} \hat h_C^{\ B} +\cdots,  
\eeqa
where the space-time indices are raised or lowered by the VEV of the metric so that $\hat h^{AB} = g^{AC} \hat h_{CD} g^{DB}$ and $\hat h_C^{\ B} = \hat h_{CD} g^{DB}$.

We then expand the Levi-Civita connection as
\beqa
\hat{\Gamma}^A_{BC} =\Gamma^A_{BC} + \dot\Gamma{}^A_{BC} +
\ddot \Gamma{}^A_{BC}+\cdots,
\eeqa
where
\beqa
\dot\Gamma^A_{BC} 
 &=&\frac{1}{2}g^{AD}\left(\hat h_{D\{B;C\}} -\hat h_{CB;D}\right), \quad
 \hat h_{AB;C} := \hat h_{AB,C} -  \Gamma^D_{CA} \hat h_{DB} -  \Gamma^D_{CB} \hat h_{AD} .
 \label{eq:Gamma1}
\eeqa
Similarly we have
\beqa
\ddot\Gamma^A_{BC} 
&=& - \frac{1}{2} \hat h^{AD}\left( \hat h_{D\{B;C\}} -  \hat h_{CB;D}\right) .
\label{eq:Gamma2}
\eeqa

Riemann curvatures are expanded as
\beqa
\hat R^A_{\ BCD} &=& R^A_{\ BCD} +\dot R{}^A_{\ BCD} +\ddot R{}^A_{\ BCD}+\cdots,
\\
\hat R_{AB} &=& R_{AB} + \dot R_{AB}+ \ddot R_{AB} +\cdots, \\
\hat R &=& R + \dot R+ \ddot R +\cdots,
\eeqa
where 
\beqa
R^A_{\ BCD} &=& \Gamma^A_{B[D,C]} + \Gamma^A_{E [C}\Gamma^E_{BD]}, \\
\dot R{}^A_{\ BCD} &=& \dot \Gamma {}^A_{B[D,C]} +  \dot \Gamma {}^A_{E [C}\Gamma^E_{BD]}
+ \Gamma^A_{E [C} \dot \Gamma {}^E_{BD]}=\dot \Gamma {}^A_{B[D;C]}, \\
\ddot R{}^A_{\ BCD} 
&=&\ddot \Gamma {}^A_{B[D;C]}
+ \dot \Gamma {}^A_{E [C}
  \dot\Gamma {}^E_{BD]} , \\
 R_{AB} &=& \Gamma^C_{A[B,C]} + \Gamma^C_{E [C}\Gamma^E_{AB]}, \\
\dot R_{AB} &=&\dot \Gamma {}^C_{A[B;C]}, \\
\ddot R_{AB} &=&
\ddot \Gamma {}^C_{A[B;C]}
+ \dot \Gamma {}^C_{E [C}\dot \Gamma {}^E_{AB]} , \\
R &=& g^{AB} R_{AB}, \\
\dot R &=& g^{AB} \dot R_{AB} -\hat h^{AB} R_{AB}, \\
\ddot R &=& g^{AB} \ddot R_{AB} -\hat h^{AB} \dot R_{AB} + \hat h^{AC} \hat h_C^{\ B} R_{AB}. 
\eeqa

Using eq.s~(\ref{eq:Gamma1}) and (\ref{eq:Gamma2}), we have
\beqa
\dot R_{AB} 
&=&\frac{1}{2} \left( \hat h^C_{\ \{A;B\}C}-\hat h_{BA}{}^{;C}_{\ \ C}-\hat h^C_{\ \ C}{}_{;AB}\right), \\
\ddot R_{AB} &=&\frac{1}{4}g^{CE} g^{FD}\left\{- 2\left[\hat h_{EF}
\left(\hat h_{D\{A;B\}} - \hat h_{BA;D}\right) \right]_{;C}
+  2\left[\hat h_{EF}\left(\hat h_{D\{A;C\}} - \hat h_{CA;D}\right) \right]_{;B} \right.
\nn 
&+&\left. \hat h_{EC;D}\left(\hat h_{F\{A;B\}} -\hat h_{BA;F}\right) 
-(\hat h_{E\{D;B\}} - \hat h_{BD;E}) (\hat h_{F\{A;C\}} - \hat h_{CA;F})
\right\}.
\eeqa

At the next to leading order, the Einstein tensor is evaluated as
\beqa
\langle \ddot G_{AB}\rangle &=& 
\langle \ddot R_{AB}\rangle 
-\frac{1}{2} \langle \hat h_{AB} \dot R \rangle -\frac{1}{2} g_{AB} \langle \ddot R\rangle, 
\eeqa
where
\beqa
 \langle \hat h_{AB} \dot R \rangle &=&  
 \langle \hat h_{AB} \{ g^{GH}  g^{CF} ( \hat h_{F G;HC}-\hat h_{HG;FC}  )  -\hat h^{CD} R_{CD }  \} \rangle, \nn
 \langle \ddot R\rangle &=& 
g^{AB} \langle \ddot R_{AB}\rangle -\frac{1}{2} g^{AE} g^{BF} g^{CD} \nn
&\times&
\left\{
 \left\langle \hat h_{EF}\left( \hat h_{D\{A;B\} C} - \hat h_{BA;DC} -\hat h_{DC;AB}\right)\rangle
 - 2 R_{AB} \langle \hat h_{ED} \hat h_{CF}\right\rangle \right\}. \notag
\eeqa

\section{2-point functions of the metric fluctuation}
\label{app:2pt}

In this appendix, we calculate various 2-point functions of $\hat h_{AB}$. 
\subsection{Preparation}
We define
\beqa
G(z_i,z_j) &:=& L^2 f (\tau_j,\tau_j) F \left(\frac{ 2 d(x_i-x_j)^2}{\tau_i+\tau_j}\right), \\
f (x,y) &=&  g(x,y)^\Delta,\ g(x,y):= \frac{2xy}{x^2+y^2}, 
\eeqa
where
\beqa
F (0)&=&1, \quad F^\prime(0) = -\frac{\Delta}{2 d}, \quad F^{\prime\prime}(0) = \frac{\Delta (\Delta+1)}{4d(d+2)}.
\eeqa
Derivatives of $f$ at $x=y$ are given by
\beqa
f_x (x,x)&=& f_y (x,x)= 0, \quad 
f_{xx}(x,x) = -\frac{\Delta}{x^2}, \quad   f_{xy} (x,x)=\frac{\Delta}{x^2}, \\
f_{xxx}(x,x) &=& \frac{3\Delta}{x^3}, \quad  f_{xxy}(x,x) = -\frac{\Delta}{x^3}, \\
f_{xxyy} (x,x) &=& \frac{3\Delta(\Delta+1)}{x^4}, \quad
f_{xxxy} (x,x)= -\frac{3\rho^2}{x^4},
\eeqa
where subscripts $x$ and $y$ denote derivatives with respect to these variables.

\subsection{The connected part of propagators}
The simplest one can be calculated from $G$ as follows.
\beqa
\langle \hat h_{AB} \hat h_{CD} \rangle &=& \langle \hat g_{AB} \hat g_{CD} \rangle_c 
=\left.  \frac{1}{n}\partial_A^1 \partial_B^2 \partial_C^3 \partial_D^4 G(z_1,z_3) G(z_2,z_4)\right\vert_{z_i=z} +  (C\leftrightarrow D) \nn 
&=& \frac{1}{n}
\frac{L^4\Delta^2}{ \tau^4}\left(\delta_{AC}\delta_{BD} +\delta_{AD}\delta_{BC}\right), 
\eeqa
where
\beqa
\left.\partial_A^1\partial_C^3 G(z_1,z_3)\right\vert_{z_1=z_3} &=& \frac{L^2\Delta}{\tau^2}\delta_{AC}.
\eeqa

Similarly 
\beqa
\langle \hat h_{AB,E} \hat h_{CD} \rangle &=&\left.  \langle  \hat g_{AB,E} \hat g_{CD} \rangle_c = \frac{1}{n}  \partial_E^1\partial_A^1\partial_C^2 G_\Delta(z_1,z_2)  \partial_B^1\partial_D^2 G_\Delta(z_1,z_2) \right\vert_{z_1=z_2=z} \nn 
&+& ( C \leftrightarrow D) +   ( A \leftrightarrow B)  +( \{A,C\} \leftrightarrow \{B,D\} ), 
\eeqa
\beqa
\langle \hat h_{AB,E} \hat h_{CD,F} \rangle &=&\left. \frac{1}{n}
\partial_E^1\partial_A^1\partial_C^2 G(z_1,z_2) \partial_F^1\partial_D^1\partial_B^2 G(z_1,z_2) \right\vert_{z_1=z_2=z} \nn 
&+& \left. \frac{1}{n}
\partial_E^1\partial_F^2\partial_A^1\partial_C^2 G(z_1,z_2) \partial_D^1\partial_B^2 G(z_1,z_2) \right\vert_{z_1=z_2=z} \nn 
&+& ( C \leftrightarrow D) +   ( A \leftrightarrow B)  +( \{A,C\} \leftrightarrow \{B,D\} ), 
\eeqa
\beqa
\langle \hat h_{AB,EF} \hat h_{CD} \rangle &=&\left. \frac{1}{n}
\partial_E^1\partial_A^1\partial_C^2 G(z_1,z_2) \partial_F^1\partial_B^1\partial_D^2 G(z_1,z_2) \right\vert_{z_1=z_2=z} \nn 
&+& \left. \frac{1}{n}
\partial_E^1\partial_F^1\partial_A^1\partial_C^2 G(z_1,z_2) \partial_B^1\partial_D^2 G(z_1,z_2) \right\vert_{z_1=z_2=z} \nn 
&+& ( C \leftrightarrow D) +   ( A \leftrightarrow B)  +( \{A,C\} \leftrightarrow \{B,D\} ) .
\eeqa

We now evaluate derivatives of $G$ as
\beqa
\left. \partial_E^1 \partial_A^1\partial_C^2 G(z_1,z_2) \right\vert_{z_1=z_2=z} &=&
-\frac{L^2\Delta}{\tau^3}\rho^{C\tau}_{AE} ,
\eeqa
\beqa
\left.  \partial_C^1\partial_D^2  \partial_A^1\partial_B^2 G(z_1,z_2)   \right\vert_{z_{1,2}=z} &=& \frac{L^2\Delta}{\tau^4}\left[3(\Delta+1)\delta^\tau_{ABCD}
+(\Delta-1)\left(\delta^\tau_{AC}\delta^d_{BD} +\delta^\tau_{BD}\delta^d_{AC}\right) 
\right. \nn 
&+& (\Delta+2)\left(\delta^\tau_{AB}\delta^d_{CD} +\delta^\tau_{AD}\delta^d_{BC} + \delta^\tau_{CD}\delta^d_{AB}+\delta^\tau_{BC}\delta^d_{AD} \right)  \nn 
&+&\left.
\frac{d}{d+2}(\Delta+1)\left( \delta^d_{AB}\delta^d_{CD} +\delta^d_{AD}\delta^d_{BC} +\delta^d_{AC}\delta^d_{BD}\right)
\right] \\
&=&  \frac{L^2\Delta}{(d+2) \tau^4}\left[ (\Delta+1) ( d\, \rho_{ABCD} -6\delta^\tau_{ABCD} )
\right. \nn 
&+& (2\Delta+d+4)\left(\delta^\tau_{AB}\delta_{CD} +\delta^\tau_{AD}\delta_{BC} + \delta^\tau_{CD}\delta_{AB}+\delta^\tau_{BC}\delta_{AD} \right)  \nn 
&+&\left. (2\Delta-2d-2)\left(\delta^\tau_{AC}\delta_{BD} +\delta^\tau_{BD}\delta_{AC}\right) \right] 
\eeqa
with $\delta^\tau_{AB}:=\delta_{A\tau}\delta_{B\tau}$,
$\delta^d_{AB} :=\delta_{AB}-\delta_{AB}^\tau$, 
$\rho_{ABCD} :=\delta_{AB}\delta_{CD}+\delta_{AC}\delta_{BD}+\delta_{AD}\delta_{BC}$, $\delta^\tau_{ABCD}:=\delta_{AB}^\tau\delta_{CD}^\tau$, and
\beqa
\left.  \partial_C^1\partial_D^1  \partial_A^1\partial_B^2 G(z_1,z_2)   \right\vert_{z_{1,2}=z} &=& -\frac{L^2\Delta}{\tau^4}\left[3\Delta\delta^\tau_{ABCD}
+(\Delta-1)\left(\delta^\tau_{AC}\delta^d_{BD} +\delta^\tau_{AD}\delta^d_{BC}+\delta^\tau_{CD}\delta^d_{AB}\right) 
\right. \nn 
&+& (\Delta+2)\left(\delta^\tau_{AB}\delta^d_{CD} +\delta^\tau_{BC}\delta^d_{AD} + \delta^\tau_{BD}\delta^d_{AC} \right)  \nn 
&+&\left.
\frac{d}{d+2}(\Delta+1)\left( \delta^d_{AB}\delta^d_{CD} +\delta^d_{AD}\delta^d_{BC} +\delta^d_{AC}\delta^d_{BD}\right)
\right] \\
&=& - \frac{L^2\Delta}{(d+2) \tau^4}\left[ (\Delta+1) ( d\, \delta_{ABCD} -6\delta^\tau_{ABCD} )
\right. \nn 
&+& (2\Delta+d+4)\left(\delta^\tau_{AB}\delta_{CD} +\delta^\tau_{BC}\delta_{AD} + \delta^\tau_{BD}\delta_{AC} \right)  \nn 
&+&\left. (2\Delta-2d-2)\left(\delta^\tau_{CD}\delta_{AB}+\delta^\tau_{AD}\delta_{BC}+\delta^\tau_{AC}\delta_{BD} \right) \right] .
\eeqa

Combining above results, we finally obtain
\beqa
\langle \hat h_{AB}  \hat h_{CD} \rangle &=& \frac{1}{n}\frac{L^4\Delta^2}{\tau^4}
\left(\delta_{AC}\delta_{BD} + \delta_{AD}\delta_{BC} \right), \\
\langle \hat h_{AB;E}  \hat h_{CD} \rangle &=& 0, \\
\langle \hat h_{AB;E}  \hat h_{CD;F} \rangle & =& - \langle \hat h_{AB;EF}  \hat h_{CD} \rangle
=  \frac{1}{n}\frac{L^4\Delta^2}{(d+2)\tau^6} [\chi_{AECF}\delta_{BD} 
+\chi_{AEDF}\delta_{BC}  \nn 
&+&  \chi_{BECF}\delta_{AD} + \chi_{BEDF}\delta_{AC} ], 
\eeqa 
where
\beqa
\chi_{AECF}& :=& 2(\Delta+1) \left\{ \frac{d}{2} \rho_{AECF}  -3\delta^\tau_{AECF} 
+  \rho_{AECF}^\tau \right\} -(d+2) \delta_{AE}\delta_{CF}, \\
\rho_{AECF} &:=& \delta_{AE}\delta_{CF}  +
\delta_{AC}\delta_{EF}  +\delta_{AF}\delta_{CE} , \\
\rho^\tau_{AECF} &:=& \delta_{AE}^\tau\delta_{CF} + \delta_{AE}\delta^\tau_{CF} +
\delta_{AC}^\tau\delta_{EF} + \delta_{AC}\delta^\tau_{EF} +\delta_{AF}^\tau\delta_{CE} + \delta_{AF}\delta^\tau_{CE}. 
\eeqa

\bibliographystyle{utphys}
\bibliography{EMTensor}

\providecommand{\href}[2]{#2}\begingroup\raggedright\begin{thebibliography}{10}

\bibitem{Maldacena:1997re}
J.~M. Maldacena, ``{The large N limit of superconformal field theories and
  supergravity},'' {\em Adv. Theor. Math. Phys.} {\bf 2} (1998) 231--252,
\href{http://arXiv.org/abs/hep-th/9711200}{{\tt hep-th/9711200}}.

\bibitem{Aharony:1999ti}
O.~Aharony, S.~S. Gubser, J.~M. Maldacena, H.~Ooguri, and Y.~Oz, ``{Large N
  field theories, string theory and gravity},'' {\em Phys. Rept.} {\bf 323}
  (2000) 183--386,
\href{http://arXiv.org/abs/hep-th/9905111}{{\tt hep-th/9905111}}.

\bibitem{Klebanov:2000me}
I.~R. Klebanov, ``{TASI lectures: Introduction to the AdS / CFT
  correspondence},'' in {\em {Strings, branes and gravity. Proceedings,
  Theoretical Advanced Study Institute, TASI'99, Boulder, USA, May 31-June 25,
  1999}}, pp.~615--650.
\newblock 2000.
\newblock
\href{http://arXiv.org/abs/hep-th/0009139}{{\tt hep-th/0009139}}.
\newblock

\bibitem{DHoker:2002nbb}
E.~D'Hoker and D.~Z. Freedman, ``{Supersymmetric gauge theories and the AdS /
  CFT correspondence},'' in {\em {Strings, Branes and Extra Dimensions: TASI
  2001: Proceedings}}, pp.~3--158.
\newblock 2002.
\newblock
\href{http://arXiv.org/abs/hep-th/0201253}{{\tt hep-th/0201253}}.
\newblock

\bibitem{tHooft:1993dmi}
G.~'t~Hooft, ``{Dimensional reduction in quantum gravity},'' in {\em {Salamfest
  1993:0284-296}}, pp.~0284--296.
\newblock 1993.
\newblock
\href{http://arXiv.org/abs/gr-qc/9310026}{{\tt gr-qc/9310026}}.
\newblock

\bibitem{Susskind:1994vu}
L.~Susskind, ``{The World as a hologram},'' {\em J. Math. Phys.} {\bf 36}
  (1995) 6377--6396,
\href{http://arXiv.org/abs/hep-th/9409089}{{\tt hep-th/9409089}}.

\bibitem{Gubser:1998bc}
S.~S. Gubser, I.~R. Klebanov, and A.~M. Polyakov, ``{Gauge theory correlators
  from non-critical string theory},'' {\em Phys. Lett.} {\bf B428} (1998)
  105--114,
\href{http://arXiv.org/abs/hep-th/9802109}{{\tt hep-th/9802109}}.

\bibitem{Witten:1998qj}
E.~Witten, ``{Anti-de Sitter space and holography},'' {\em Adv. Theor. Math.
  Phys.} {\bf 2} (1998) 253--291,
\href{http://arXiv.org/abs/hep-th/9802150}{{\tt hep-th/9802150}}.

\bibitem{Banks:1998dd}
T.~Banks, M.~R. Douglas, G.~T. Horowitz, and E.~J. Martinec, ``{AdS dynamics
  from conformal field theory},''
\href{http://arXiv.org/abs/hep-th/9808016}{{\tt hep-th/9808016}}.

\bibitem{Balasubramanian:1998sn}
V.~Balasubramanian, P.~Kraus, and A.~E. Lawrence, ``{Bulk versus boundary
  dynamics in anti-de Sitter space-time},'' {\em Phys. Rev.} {\bf D59} (1999)
  046003,
\href{http://arXiv.org/abs/hep-th/9805171}{{\tt hep-th/9805171}}.

\bibitem{Balasubramanian:1998de}
V.~Balasubramanian, P.~Kraus, A.~E. Lawrence, and S.~P. Trivedi, ``{Holographic
  probes of anti-de Sitter space-times},'' {\em Phys. Rev.} {\bf D59} (1999)
  104021,
\href{http://arXiv.org/abs/hep-th/9808017}{{\tt hep-th/9808017}}.

\bibitem{Balasubramanian:1999ri}
V.~Balasubramanian, S.~B. Giddings, and A.~E. Lawrence, ``{What do CFTs tell us
  about Anti-de Sitter space-times?},'' {\em JHEP} {\bf 03} (1999) 001,
\href{http://arXiv.org/abs/hep-th/9902052}{{\tt hep-th/9902052}}.

\bibitem{Bena:1999jv}
I.~Bena, ``{On the construction of local fields in the bulk of AdS(5) and other
  spaces},'' {\em Phys. Rev.} {\bf D62} (2000) 066007,
\href{http://arXiv.org/abs/hep-th/9905186}{{\tt hep-th/9905186}}.

\bibitem{Lifschytz:2000bj}
G.~Lifschytz and V.~Periwal, ``{Schwinger-Dyson = Wheeler-DeWitt: Gauge theory
  observables as bulk operators},'' {\em JHEP} {\bf 04} (2000) 026,
\href{http://arXiv.org/abs/hep-th/0003179}{{\tt hep-th/0003179}}.

\bibitem{Nozaki:2013vta}
M.~Nozaki, T.~Numasawa, A.~Prudenziati, and T.~Takayanagi, ``{Dynamics of
  Entanglement Entropy from Einstein Equation},'' {\em Phys. Rev.} {\bf D88}
  (2013), no.~2, 026012,
\href{http://arXiv.org/abs/1304.7100}{{\tt 1304.7100}}.

\bibitem{Bhattacharya:2013bna}
J.~Bhattacharya and T.~Takayanagi, ``{Entropic Counterpart of Perturbative
  Einstein Equation},'' {\em JHEP} {\bf 10} (2013) 219,
\href{http://arXiv.org/abs/1308.3792}{{\tt 1308.3792}}.

\bibitem{Faulkner:2013ica}
T.~Faulkner, M.~Guica, T.~Hartman, R.~C. Myers, and M.~Van~Raamsdonk,
  ``{Gravitation from Entanglement in Holographic CFTs},'' {\em JHEP} {\bf 03}
  (2014) 051,
\href{http://arXiv.org/abs/1312.7856}{{\tt 1312.7856}}.

\bibitem{Faulkner:2017tkh}
T.~Faulkner, F.~M. Haehl, E.~Hijano, O.~Parrikar, C.~Rabideau, and
  M.~Van~Raamsdonk, ``{Nonlinear Gravity from Entanglement in Conformal Field
  Theories},'' {\em JHEP} {\bf 08} (2017) 057,
\href{http://arXiv.org/abs/1705.03026}{{\tt 1705.03026}}.

\bibitem{Hollands:2012sf}
S.~Hollands and R.~M. Wald, ``{Stability of Black Holes and Black Branes},''
  {\em Commun. Math. Phys.} {\bf 321} (2013) 629--680,
\href{http://arXiv.org/abs/1201.0463}{{\tt 1201.0463}}.

\bibitem{Sarosi:2017rsq}
G.~S\'arosi and T.~Ugajin, ``{Modular Hamiltonians of excited states, OPE
  blocks and emergent bulk fields},''
\href{http://arXiv.org/abs/1705.01486}{{\tt 1705.01486}}.

\bibitem{Osborn:1991gm}
H.~Osborn, ``{Weyl consistency conditions and a local renormalization group
  equation for general renormalizable field theories},'' {\em Nucl. Phys.} {\bf
  B363} (1991)
486--526.

\bibitem{Imbimbo:1999bj}
C.~Imbimbo, A.~Schwimmer, S.~Theisen, and S.~Yankielowicz, ``{Diffeomorphisms
  and holographic anomalies},'' {\em Class. Quant. Grav.} {\bf 17} (2000)
  1129--1138,
\href{http://arXiv.org/abs/hep-th/9910267}{{\tt hep-th/9910267}}.

\bibitem{Papadimitriou:2004ap}
I.~Papadimitriou and K.~Skenderis, ``{AdS / CFT correspondence and geometry},''
  {\em IRMA Lect. Math. Theor. Phys.} {\bf 8} (2005) 73--101,
\href{http://arXiv.org/abs/hep-th/0404176}{{\tt hep-th/0404176}}.

\bibitem{Lee:2012xba}
S.-S. Lee, ``{Background independent holographic description : From matrix
  field theory to quantum gravity},'' {\em JHEP} {\bf 10} (2012) 160,
\href{http://arXiv.org/abs/1204.1780}{{\tt 1204.1780}}.

\bibitem{Lee:2013dln}
S.-S. Lee, ``{Quantum Renormalization Group and Holography},'' {\em JHEP} {\bf
  01} (2014) 076,
\href{http://arXiv.org/abs/1305.3908}{{\tt 1305.3908}}.

\bibitem{Skenderis:2002wp}
K.~Skenderis, ``{Lecture notes on holographic renormalization},'' {\em Class.
  Quant. Grav.} {\bf 19} (2002) 5849--5876,
\href{http://arXiv.org/abs/hep-th/0209067}{{\tt hep-th/0209067}}.

\bibitem{Shyam:2016zuk}
V.~Shyam, ``{General Covariance from the Quantum Renormalization Group},'' {\em
  Phys. Rev.} {\bf D95} (2017), no.~6, 066003,
\href{http://arXiv.org/abs/1611.05315}{{\tt 1611.05315}}.

\bibitem{Shyam:2017qlr}
V.~Shyam, ``{Connecting Holographic Wess--Zumino Consistency Conditions to the
  Holographic Anomaly},''
\href{http://arXiv.org/abs/1712.07955}{{\tt 1712.07955}}.

\bibitem{Aoki:2015dla}
S.~Aoki, K.~Kikuchi, and T.~Onogi, ``{Geometries from field theories},'' {\em
  PTEP} {\bf 2015} (2015), no.~10, 101B01,
\href{http://arXiv.org/abs/1505.00131}{{\tt 1505.00131}}.

\bibitem{Aoki:2016ohw}
S.~Aoki, J.~Balog, T.~Onogi, and P.~Weisz, ``{Flow equation for the large $N$
  scalar model and induced geometries},'' {\em PTEP} {\bf 2016} (2016), no.~8,
  083B04,
\href{http://arXiv.org/abs/1605.02413}{{\tt 1605.02413}}.

\bibitem{Aoki:2017bru}
S.~Aoki and S.~Yokoyama, ``{Flow equation, conformal symmetry, and anti-de
  Sitter geometry},'' {\em PTEP} {\bf 2018} (2018), no.~3, 031B01,
\href{http://arXiv.org/abs/1707.03982}{{\tt 1707.03982}}.

\bibitem{Aoki:2017uce}
S.~Aoki and S.~Yokoyama, ``{AdS geometry from CFT on a general conformally flat
  manifold},'' {\em Nucl. Phys.} {\bf B933} (2018) 262--274,
\href{http://arXiv.org/abs/1709.07281}{{\tt 1709.07281}}.

\bibitem{Albanese:1987ds}
{\bf APE} Collaboration, M.~Albanese {\em et al.}, ``{Glueball Masses and
  String Tension in Lattice QCD},'' {\em Phys. Lett.} {\bf B192} (1987)
163--169.

\bibitem{Narayanan:2006rf}
R.~Narayanan and H.~Neuberger, ``{Infinite N phase transitions in continuum
  Wilson loop operators},'' {\em JHEP} {\bf 03} (2006) 064,
\href{http://arXiv.org/abs/hep-th/0601210}{{\tt hep-th/0601210}}.

\bibitem{Luscher:2010iy}
M.~Luscher, ``{Properties and uses of the Wilson flow in lattice QCD},'' {\em
  JHEP} {\bf 08} (2010) 071, \href{http://arXiv.org/abs/1006.4518}{{\tt
  1006.4518}}.
[Erratum: JHEP03,092(2014)].

\bibitem{Jevicki:1998qs}
A.~Jevicki, Y.~Kazama, and T.~Yoneya, ``{Quantum metamorphosis of conformal
  transformation in D3-brane Yang-Mills theory},'' {\em Phys. Rev. Lett.} {\bf
  81} (1998) 5072--5075,
\href{http://arXiv.org/abs/hep-th/9808039}{{\tt hep-th/9808039}}.

\bibitem{Das:2003vw}
S.~R. Das and A.~Jevicki, ``{Large N collective fields and holography},'' {\em
  Phys. Rev.} {\bf D68} (2003) 044011,
\href{http://arXiv.org/abs/hep-th/0304093}{{\tt hep-th/0304093}}.

\bibitem{Klebanov:2002ja}
I.~Klebanov and A.~Polyakov, ``{AdS dual of the critical O(N) vector model},''
  {\em Phys.Lett.} {\bf B550} (2002) 213--219,
\href{http://arXiv.org/abs/hep-th/0210114}{{\tt hep-th/0210114}}.

\bibitem{Giombi:2013fka}
S.~Giombi and I.~R. Klebanov, ``{One Loop Tests of Higher Spin AdS/CFT},'' {\em
  JHEP} {\bf 12} (2013) 068,
\href{http://arXiv.org/abs/1308.2337}{{\tt 1308.2337}}.

\bibitem{Giombi:2014iua}
S.~Giombi, I.~R. Klebanov, and B.~R. Safdi, ``{Higher Spin AdS$_{d+1}$/CFT$_d$
  at One Loop},'' {\em Phys. Rev.} {\bf D89} (2014), no.~8, 084004,
\href{http://arXiv.org/abs/1401.0825}{{\tt 1401.0825}}.

\bibitem{Camporesi:1993mz}
R.~Camporesi and A.~Higuchi, ``{Arbitrary spin effective potentials in anti-de
  Sitter space-time},'' {\em Phys. Rev.} {\bf D47} (1993)
3339--3344.

\bibitem{Giombi:2014yra}
S.~Giombi, I.~R. Klebanov, and A.~A. Tseytlin, ``{Partition Functions and
  Casimir Energies in Higher Spin AdS$_{d+1}$/CFT$_d$},'' {\em Phys. Rev.} {\bf
  D90} (2014), no.~2, 024048,
\href{http://arXiv.org/abs/1402.5396}{{\tt 1402.5396}}.

\bibitem{Jevicki:2014mfa}
A.~Jevicki, K.~Jin, and J.~Yoon, ``{1/N and loop corrections in higher spin
  AdS$_4$/CFT$_3$ duality},'' {\em Phys. Rev.} {\bf D89} (2014), no.~8, 085039,
\href{http://arXiv.org/abs/1401.3318}{{\tt 1401.3318}}.

\bibitem{Beccaria:2014xda}
M.~Beccaria and A.~A. Tseytlin, ``{Higher spins in AdS$_{5}$ at one loop:
  vacuum energy, boundary conformal anomalies and AdS/CFT},'' {\em JHEP} {\bf
  11} (2014) 114,
\href{http://arXiv.org/abs/1410.3273}{{\tt 1410.3273}}.

\bibitem{Bae:2016rgm}
J.-B. Bae, E.~Joung, and S.~Lal, ``{One-loop test of free SU(N ) adjoint model
  holography},'' {\em JHEP} {\bf 04} (2016) 061,
\href{http://arXiv.org/abs/1603.05387}{{\tt 1603.05387}}.

\bibitem{Aoki:2016env}
S.~Aoki, J.~Balog, T.~Onogi, and P.~Weisz, ``{Flow equation for the scalar
  model in the large $N$ expansion and its applications},'' {\em PTEP} {\bf
  2017} (2017), no.~4, 043B01,
\href{http://arXiv.org/abs/1701.00046}{{\tt 1701.00046}}.

\end{thebibliography}\endgroup

\end{document}